\documentclass[12pt,amstex,aps,amsfonts,prsty]{article} 
\usepackage[dvips]{graphicx}
\usepackage{a4wide}
\renewcommand{\baselinestretch}{1.1} 
\begin{document}
\begin{center}
{\bf Propagation Dynamics of a Particle Phase in a Single-File Pore.}
\end{center}
{\bf A.M.Lacasta$^{1}$, J.M.Sancho$^2$, F.Sagues$^3$ and  G.Oshanin$^4$\\}
{$^1$ Departament de Fisica Aplicada, Universitat Polit\`ecnica de
Catalunya, E-08028 Barcelona, Spain\\
}
{$^2$ Departament d'Estructura i Constituents de la Mat\`eria, Universitat
de Barcelona,  E-08028 Barcelona, Spain\\
}
{$^3$  Departament de Quimica Fisica, Universitat
de Barcelona,  E-08028 Barcelona, Spain\\
}
{$^4$ Laboratoire de Physique Th{\'e}orique des Liquides, 
Universit{\'e} Paris 6, 75252 Paris, France\\
}

{\Large \bf Abstract} 

\vspace{0.2in}

We study propagation dynamics of a particle phase in a single-file pore connected to a
reservoir of particles (bulk liquid phase). We show that the total mass $M(t)$ 
of particles
entering the pore up to time $t$ grows as $M(t) = 2 m(J,\rho_F) \sqrt{D_0 t}$, where $D_0$ is
the "bare" diffusion coefficient and the prefactor $m(J,\rho_F)$ is a non-trivial function of
the reservoir density $\rho_F$ and the amplitude $J$ of attractive particle-particle interactions.
Behavior of the dynamic density profiles is also discussed.

\vspace{0.3in}

{\Large \bf Introduction} 

\vspace{0.2in}

Particles transport across microscopic pores is an important step 
in a vast variety of
biological, chemical engineering and industrial processes, including  
drug release, catalyst preparation and operation, 
separation technologies, especially biological and biochemical, 
tertiary oil recovery, drying and chromatography \cite{1,2,3}. 

Man-made or naturally occuring porous materials contain a wide range 
of pore sizes, from meso- to micro- or even nanoscales, in which case
the pore diameter is comparable to the molecular size. 
Such molecular sized channels, of order of a few
Angstroms only, appear, for instance, in biological membranes, 
and are specific to water and ion transport which participate in
hydrostatic or osmotic pressure controlled cellular volume 
regulation \cite{alberts}.
Carbon nanotubes or zeolites, such as, Mordenite, L,
AIPO$_4$-5, ZSM-12, may also contain many channels of nearly
 molecular diameter and can
selectively absorb fluids serving as remarkable molecular sieves \cite{meier}. 
 For
example, AIPO$_4$-5 is composed of nonintersecting and 
approximately cylindrical pores of nominal
diameter $7.3$ Angstroms. 

\begin{figure}[ht]
\begin{center}
\includegraphics*[scale=0.45]{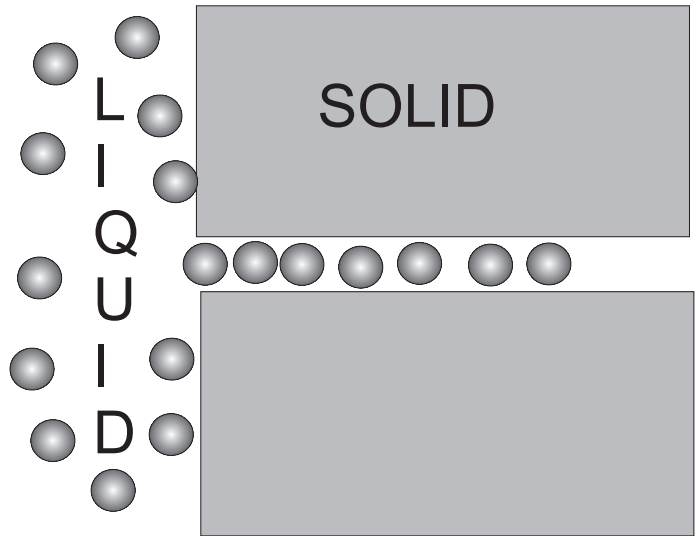}
\caption{\label{Fig1} {\small  Solid with a single-file pore in contact with a liquid phase.}}
\end{center}
\end{figure} 

A salient feature of transport in molecularly sized pores is 
that there is a dramatic difference
between the diffusion of adsorbates whose size is much smaller than the pore diameter, and
those whose size is comparable to it. If the diameter of the diffusing guest molecules
exceeds the pore radii, but is still less than its diameter,
the particles are able to enter the pore but not able to bypass each other such that initial given order is
striktly maintained (see Fig.1). Here, diffusion obeys the so-called "single-file" behavior; 
 the mean-square displacements of
tracer molecules do 
follow $\overline{X^2(t)} = 2 F t^{1/2}$ \cite{gupta,kukla,keffer}, 
where $F$ is referred to as
the "single-file" mobility; evidently, such a behavior is remarkably different from  the conventional
diffusive law $\overline{X^2(t)} = 2 D t$, observed in situations when the guest
molecules are able to bypass each other. Single-file diffusion 
was evidenced by pulsed field gradient NMR measurements with a variety of zeolites and
guest species \cite{gupta,kukla}; in particular, the law  
$\overline{X^2(t)} = 2 F t^{1/2}$
has been identified
experimentally for C$_2$H$_6$ \cite{gupta} and CF$_4$ \cite{hahn} in  AIPO$_4$-5.

Diffusion of absorbed particles in single-file pores 
has also been  the issue of a considerable
theoretical interest recently. Several approaches have been proposed, 
based mostly on the lattice-gas-type models,  and such properties as concentration profiles or 
steady-state particle currents have been evaluated \cite{rodenbeck,chou1,chou2}. 
As well, a great deal of Monte Carlo and Molecular
Dynamics simulations  has been devoted to the problem \cite{murad,sholl,sahimi}, providing a deeper understanding of the transport mechanisms in single-file
pores.  On the other hand, still little is known about non-stationary behavior in single-file systems; in particular, how fast do the
particle phase propagates within the single-file pores or how does the total mass (or number) 
of particles entering the pore up to
time $t$ grows with time?

In the present paper we focus on the challenging question of 
the particle phase  propagation dynamics
 in the single-file pores.
We report here some 
preliminary results; a detailed account will be published elsewhere \cite{general}. 
More specifically, we
consider a single-file pore in contact with a reservoir of particles (bulk liquid phase),
 which maintains a fixed particle density at the
entrance to the pore. 
The pore is modelled, in a usual fashion, as a
one-dimensional regular lattice whose sites support, at most, a single occupancy; 
the particles interaction potential consists of an
abrupt, hard-core repulsive part, which insures single-occupancy, and is attractive, with an amplitude $J \geq 0$,  
for the nearest-neighboring particles only. Introducing then a standard, interacting 
lattice-gas dynamic rules (see e.g. \cite{leb}), we derive evolution equation
for the local variables describing mean occupation (density) of the lattice sites, 
which is analysed both analytically and
numerically. Within our appoach, we define the evolution of the total number $M(t)$ 
of particles entering 
the single-file pore up to time
$t$ and also discuss the dynamics of the density distribution function within the
 pore. We show that the growth of 
$M(t)$ is described by $M(t) = 2 m(J,\rho_F) \sqrt{ D_0 t}$, which can be thought of
 as the microscopic analog of the Washburn
equation, where $D_0$ denotes the "bare" diffusion coefficient, while the prefactor 
$m(J,\rho_F)$ is a non-trivial function of the attractive
interactions amplitude $J$ and the density $\rho_F$ at the entrance to the pore.

\vspace{0.3in}

{\Large \bf The model}

\vspace{0.2in}

Following earlier works \cite{rodenbeck,chou1,chou2}, as well as  a conceptually close 
analysis of an upward creep dynamics of ultrathin liquid films in the capillary rise geometries \cite{bur}, which appears to be
 well-adapted to
the single-file dynamics, 
we model the single-file system under study (Fig.1)
as a  semi-infinite
linear chain of equidistantly placed sites $X$ (with spacing $\sigma$), 
attached to a reservoir of particles maintained at a 
constant chemical potential $\mu$ (see Fig.2). 
Each pair of sites is 
separated by a potential barrier of height $E_B$, which sets 
the typical time scale $\tau_B$.
Note that the spacing $\sigma$ can be defined, 
in case of hard solids, as the interwell distance  of a periodic 
potential
describing the interactions of the particles with the solid atoms and  $\tau_B$
is related to $E_B$ and the reciprocal 
temperature $\beta = 1/T$ through the Arrhenius formula.  
 For soft solids, in which case the dominant dissipation channel is due to mutual particle-particle interactions, 
$\sigma$ can be thought of 
as the typical distance travelled by
particles before successive collisions; here, $\tau_B$ is just the ballistic travel time.

\begin{figure}[ht]
\begin{center}
\includegraphics*[scale=0.7]{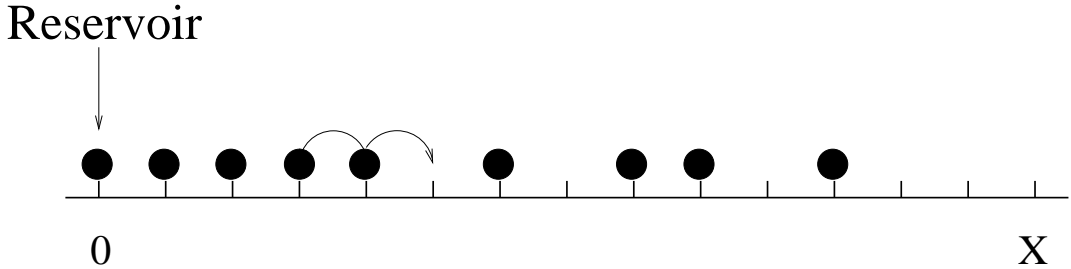}
\caption{\label{Fig2} {\small  Effective model for transport in a single-file pore.}}
\end{center}
\end{figure}

Further on, 
we define the particles interaction
potential $U(X)$ as:
 
\begin{equation}
U(X) = \left\{\begin{array}{ll}
0,  \;\;\;   \mbox{$X >  \sigma$,} \nonumber\\
- J , \;\;\;     \mbox{$X = \sigma$,} \nonumber\\
+ \infty,   \;\;\;   \mbox{$X < \sigma$,}
\end{array}
\right.
\end{equation} 
i.e. 
we suppose that the interaction potential between particle is a
hard-core exclusion, which prevents multiple occupancy of any site, and attraction with
an amplitude $J$, ($J\geq 0$), between the nearest-neighboring particles only. 
Occupation of the site $X$ at time moment $t$ for a given realization of the process
will be described 
then by the Boolean variable $\eta_t(X)$, such that

\begin{equation}
\eta_t(X) = \left\{\begin{array}{ll}
1,    \mbox{the site $X$ is occupied, } \nonumber\\
0,     \mbox{otherwise.}
\end{array}
\right.
\end{equation} 

Consequently, the 
interaction energy $U_t(X)$ of the particle occupying  at time $t$  the site
$X$ for a given realization of the dynamical process is

\begin{equation}
\label{en}
U_t(X) = - J \Big(\eta_t(X+\sigma) + \eta_t(X-\sigma)\Big).
\end{equation}

Lastly, we define the particle dynamics (see \cite{leb} for more details). We suppose that 
at
time moment $t$ any particle occupying site $X$
waits an exponential time with mean $\tau_B$ 
and then selects a jump direction with the
probabilty

\begin{equation}
\label{rate}
p(X|X') = Z^{-1} \exp\Big[ \frac{\beta}{2} \Big(U_t(X) - U_t(X')\Big)\Big],
\; \; \; \sum_{X'} p(X|X') = 1,
\end{equation}
where $Z$ is the normalization,  $X' (= X \pm \sigma)$ denotes here the target neighboring 
site and the sum over $X'$ means
the sum over all nearest neighbors of the site $X$. As soon as the target site is chosen, the particle attempts
to hop onto it; the hop is instantaneously fulfilled if the target site is empty;
otherwise, the particle remains at its position. Physically, it means that repulsive interactions
are very short-ranged - much shorter than the lattice spacing, and particles "learn" about them
only when they attempt to land onto some already occupied site. In turn, attractive
interactions are felt within the distance equal to the lattice spacing and hence, influence the
choice of the jump direction in order to minimize the total energy of the system.

\vspace{0.3in}

{\Large \bf Evolution equations}

\vspace{0.2in}

Now, let $ \rho_t(X) = \overline{\eta_t(X)}$, where the overbar denotes averaging with respect to different 
realizations of the process. Assuming local equilibrium, we find then that the time evolution of $ \rho_t(X)$ 
is governed by the following balance equation \cite{general}:

\begin{eqnarray}
\label{l}
\tau_B \dot{\rho_{t}}(X) &=& \Big(1 - \rho_{t}(X)\Big)
\Big[\rho_{t}(X-\sigma) \overline{p(X-\sigma|\sigma)} + \rho_{t}(X+\sigma) \overline{p(X+\sigma|X)}\Big] - \nonumber\\
&-& \rho_{t}(X) \Big[(1 - \rho_{t}(X+\sigma)) \overline{p(X|X+\sigma)} + (1 - \rho_{t}(X-\sigma)) \overline{p(X|X-\sigma)}\Big],
\end{eqnarray}
where the average transition rate $\overline{p(X|X')}$ obeys Eq.(\ref{rate}) with $\eta_t(X)$ replaced by $\rho_t(X)$. 
Equation  (\ref{l})
accounts for the fact that a particle may appear at time moment $t$
on an $empty$  site $X$ by hopping from the $occupied$  
sites $X \pm \sigma$  with corresponding transition probabilities dependent on the interaction
energy of the system; and may leave the $occupied$ site $X$ for $unoccupied$ sites $X \pm \sigma$.

Next, we turn to the so-called diffusion limit, assuming that $\tau_B$ scales as $\sigma^2$; we thus suppose that 
$\sigma \to 0$, $\tau_B \to 0$, 
but the ratio $\sigma^2/\tau_B =
const = 2 D_0$, where $D_0$ is the diffusion coefficient describing motion of an individual, isolated particle. 
Expanding
$\rho_{t}(X \pm \sigma)$ and $\overline{p(X|X\pm\sigma)}$ in the Taylor 
series up to the second order in powers of the lattice spacing $\sigma$, we arrive at the desired 
dynamical equation of the form 

\begin{equation}
\label{desired1}
\dot{\rho}_t(X) = D_0 \frac{\partial }{\partial X} \Big[\frac{\partial }{\partial X}
+ \beta \rho_t(X) (1- \rho_t(X)) \frac{\partial U_t(X)}{\partial X}\Big],
\end{equation}
where $U_t(X)$ is the interaction energy at point $X$ defined by Eq.(\ref{en}).

Now, a few comments on Eq.(\ref{desired1}) are in order. Note first 
that Eq.(\ref{desired1})
is a Burgers-type equation with an environment dependent force,

\begin{equation}
\label{force}
\frac{\partial U_t(X)}{\partial X} \approx - 2 J \frac{\partial \rho_t(X)}{\partial X} - \sigma^2 J 
\frac{\partial^2 U_t(X)}{\partial X^2}.
\end{equation}

When only the first term on the rhs of Eq.(\ref{force}) is taken into acount, we get  
from Eq.(\ref{desired1}) a one-dimensional diffusion equation

\begin{equation}
\label{m}
\dot{\rho}_t(X) = \frac{\partial }{\partial X} D(\rho_t(X)) \frac{\partial }{\partial X} \rho_t(X),
\end{equation}
with a field-dependent diffusion coefficient

\begin{equation}
 D(\rho_t(X)) = D_0 \Big(1 - 2 \beta J \rho_t(X) (1 - \rho_t(X))\Big),
\end{equation}
which is precisely the equation derived earlier by Lebowitz et al \cite{lebowitz} and describing 
hydrodynamic limit 
dynamics of a system of mutual interacting particles undergoing ballistic motion. 
On the other hand, if we keep the second term on the rhs of  Eq.(\ref{force}), (which is appropriate if we consider some
steady-state solutions \cite{general}), we will obtain the customary equation 
of the form  

\begin{equation}
\label{field}
\dot{\rho}_t(X) = \frac{\partial }{\partial X} M(\rho_t(X)) \frac{\partial }{\partial X} \frac{\delta 
{\cal F}(\rho_t(X))}{\delta \rho_t(X)},
\end{equation}
where the mobility $M(\rho_t(X))$ is given by

\begin{equation}
\label{mob}
M(\rho_t(X)) = \rho_t(X) (1- \rho_t(X)), 
\end{equation}
while local free energy ${\cal F}(\rho)$  obeys

\begin{equation}
{\cal F}(\rho)  = \int dX \Big(f(\rho) + \frac{\sigma^2 \beta J}{2}  \Big(\frac{\partial \rho}{\partial X}\Big)^2\Big),
\end{equation}
with

\begin{equation}
\label{j}
f(\rho) = \rho \; \ln{\rho} + (1 - \rho) \ln{(1 - \rho)} + \beta J \rho (1 - \rho).
\end{equation}

Curiously enough, $f(\rho)$, which has been derived in our work starting from
a microscopic dynamical model obeying the detailed balance condition, has exactly the same
form as the phenomenological Flory-Huggins-de Gennes local free energy density \cite{q,p}. 
Note also that for $\beta J \geq 2$,
the local free energy  $f(\rho)$ in Eq.(\ref{j}) has a double-well 
structure whose minima approach $0$ and $1$ as $\beta J$
increases. This implies that the Onsager mobility in Eq.(\ref{mob}) never reaches
negative values, contrary to the behavior 
predicted by Eq.(\ref{m}) for which one
has  $D(\rho) < 0 $ when $\rho_{c,-} < \rho <  \rho_{c,+}$, 
with
\begin{equation}
\rho_{c,\pm} = \frac{1}{2}(1 \pm \sqrt{1 - 2/\beta J}).
\end{equation}

Finally, we define the appropriate boundary conditions. As a matter of fact, 
any particle, in order to enter to the nanopore from the liquid phase, has to surmount 
an additional barrier $E_M$ related to the enthalpic energy difference between the particle 
within the pore and in the bulk liquid
phase \cite{chou1,chou2}. Supposing that the reservoir (bulk liquid phase) is in equilibrium 
with the particle phase within the single-file pore,  
we thus stipulate, following a similar analysis in \cite{bur},
 that the reservoir maintains a constant density $\rho_F$ (see \cite{chou1,chou2} for
relation between $\rho_F$ and energetic parameters) at 
the entrance of the pore (site $X = 0$ in Fig.2). 
Second boundary condition is rather evident, we just suppose
 that $\rho_t(X)$ vanishes as $X \to \infty$ at fixed $t$.
Consequently, 
Eq.(\ref{desired1}) (or Eq.(\ref{field})) is to be solved subject to the conditions 

\begin{equation}
\label{bc}
\rho_t(X = 0) = \rho_F,  \;\;\; \rho_{t}(X \to \infty) = 0.
\end{equation}

Below we discuss solutions of Eqs.(\ref{desired1}) and (\ref{field}) obeying these two boundary conditions.

\vspace{0.3in}

{\Large \bf Results}

\vspace{0.2in} 

We focus here on the time-evolution of the total mass of particles $M(t)$, having
 entered the single-file pore  
up to time $t$, i.e., 

\begin{equation}
\label{mass1}
M(t) = \int^{\infty}_{0} dX \; \rho_t(X)
\end{equation}

To define the time-dependence of $M(t)$, it is expedient to  turn
 to the scaled variable  $\omega = X/2
\sqrt{D_0 t}$. In terms of this variable Eq.(\ref{field}) attains the form

\begin{eqnarray}
\label{r}
\frac{d^2 \rho(\omega)}{d \omega^2} &+& 2 \omega \; 
\frac{d\rho(\omega)}{d \omega} - 2
\beta J \frac{d}{d\omega} \Big[\rho(\omega) \Big(1 -
\rho(\omega)\Big) \frac{d\rho(\omega)}{d \omega}  \Big]   -  \nonumber\\
&-& \beta J \Big(\frac{\sigma^2}{4 D_0 t} \Big) \frac{d}{d\omega} \Big[\rho(\omega) \Big(1 -
\rho(\omega)\Big) \frac{d^3\rho(\omega)}{d \omega^3}  \Big]   = 0,
\end{eqnarray}
while the boundary conditions in Eq.(\ref{bc})
become
$\rho(\omega = 0) = \rho_F$ and $\rho(\omega \to \infty) = 0$. 

Note now that the term in the second line on the rhs of Eq.(\ref{r}), associated with the second term in 
the expansion of the
interaction energy in the Taylor series, Eq.(\ref{force}), is irrelevant to the dynamics as $t \to \infty$, since 
it is multiplied by a vanishing function of time. Hence, the propagation dynamics can be adequately 
described by Eq.(\ref{m}), which is not
sufficient, however, for description of the steady-state characteristics, 
such as, e.g. steady-state particle current through a
finite pore \cite{general}. 

Now, in terms of the scaled variable $\omega$, the total 
mass of particles $M(t)$ reads

\begin{equation}
\label{mass}
M(t) =  2 \; m(J, \rho_F) \; \sqrt{D_0 t},
\end{equation} 

where the prefactor $m(J, \rho_F)$ is determined by

\begin{equation}
m(J, \rho_F) = \int^{\infty}_{0} d\omega \; \rho(\omega)
\end{equation}

Consequently, Eq.(\ref{mass}), which can be thought of as 
the microscopic analog of the Washburn equation, signifies that the mass of
particles grows in proportion to the square-root of time.  Note that similar result
has been obtained in \cite{bur} for $J \equiv 0$, in which case $m(J, \rho_F) =
\rho_F/\sqrt{\pi} $.

Before we discuss behavior of the prefactor $m(J, \rho_F)$, it might be instructive to
understand what are the physical processes underlying the $M(t) \sim \sqrt{t}$ behavior (see also
\cite{bur} for more detailed discussion). To do
this, let us first recollect that the boundary condition $\rho_t(X = 0) = \rho_F =
const$ is tantamount to the assumption  that the reservoir is in equilibrium with the particle phase in the
pore. Turning next to the model depicted in Fig.2, we notice that a jump of the
rightmost particle of the particle phase away of the reservoir, leads to creation of a "vacancy".
When this vacancy manages to reach diffusively, due to redistribution of particles,
the entrance of the pore, it perturbs the equilibrium and gets filled by a particle
from the reservoir. Hence, the mass of particles within the pore $M(t)$ is
proportional to the current ${\cal J}$ 
of vacancies from the front of the propagating phase\footnote{Note, however,  that there are some  
subtleties concerning propagation of the rightmost particle of the phase growing in the single-file pore. As a matter of fact,
it has been shown in \cite{bur} that in a similar model without attractive interactions (i.e. $J = 0$) the mean displacement of
the rightmost particle follows $\overline{X}(t) \sim \sqrt{t \ln{(t)}}$, i.e. grows at a faster rate than the mass $M(t)$.
Similar behavior is observed for the front of the pahse, propagating in the single-file pore, 
also in the interacting case, i.e. for arbitrary $J$ \cite{general}.}, $M(t) \sim {\cal J}$, where
the current  ${\cal J} \sim 1/L$, $L$ being the distance travelled by the rightmost particle
away from the entrance of the pore. Consequently, $M(t) \sim L \sim 1/L$, which yields
eventually the  $M(t) \sim \sqrt{t}$ law. Note also that, from the viewpoint of the
underlying physics, the dynamical process under study is intrinsically related to such
phenomena as directional solidification, freezing, limited by diffusive motion of the
latent heat, or Stefan problem.

To define the prefactor $m(J, \rho_F)$ we have solved Eq.(\ref{r}) (with the last term
set equal to zero) numerically, 
for various values of the system's parameters $\rho_F$ and $J$.
These results are summarized in Fig.3 where we depict $m(J, \rho_F)$ as a function of
$\rho_F$ for several different values of $\beta J$.    

\begin{figure}[ht]
\begin{center}
\includegraphics*[scale=0.7]{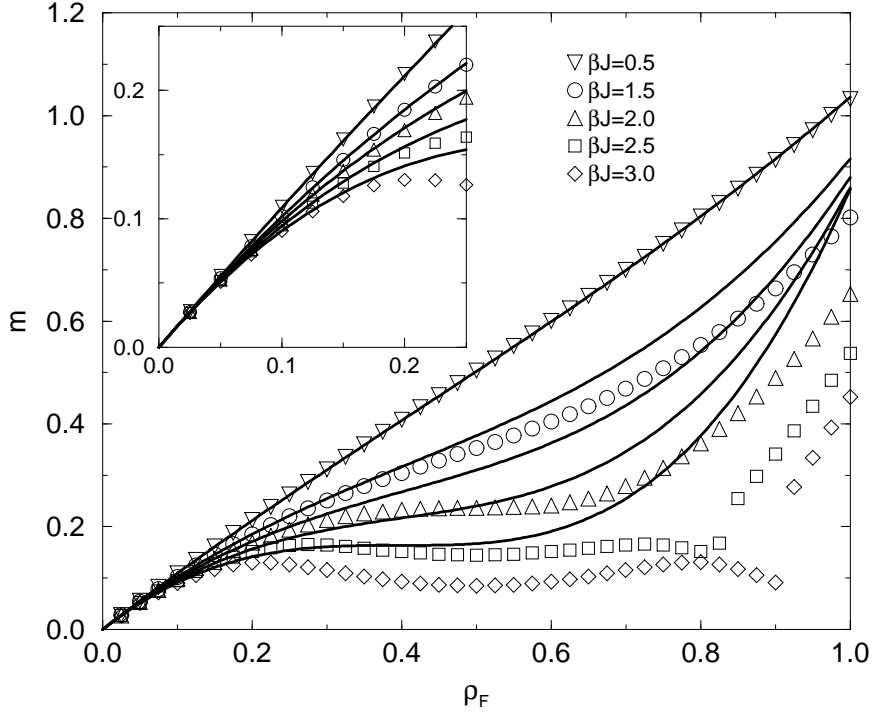}
\caption{\label{Fig3} {\small  
The prefactor $m = m(J,\rho_F)$ versus the density $\rho_F$ for different values of the amplitude
$J$ of
the attractive particle-particle interactions. Symbols denote the results of numerical solution of Eq.(\ref{r}), while the
solid lines correspond to the analytical result in Eq.(\ref{mm}). The inset displays the behavior for small $\rho_F$.}}
\end{center}
\end{figure} 

Now, one notices that for any values of $\rho_F$ and $\beta J$ the prefactor  
$m(J, \rho_F) > 0$, which implies that, as one can expect on intuitive grounds, there
is no transition in the one-dimensional system under study and the particle phase
propagates into the pore as soon as $\rho_F  > 0$. On the other hand, the prefactor 
$m(J, \rho_F)$ depends on the system's parameters in a quite non-trivial fashion.
While for relatively small $\beta J$ the prefactor $m(J,\rho_F)$ varies with $\rho_F$ almost linearly, for
large $\beta J$ some saturation effect occurs, followed by, for larger 
$\beta J$, a non-monotoneous $\rho_F$-dependence, and eventually, after a
cusp-like variation, with a rapid growth. 

\begin{figure}[ht]
\begin{center}
\includegraphics*[scale=0.8]{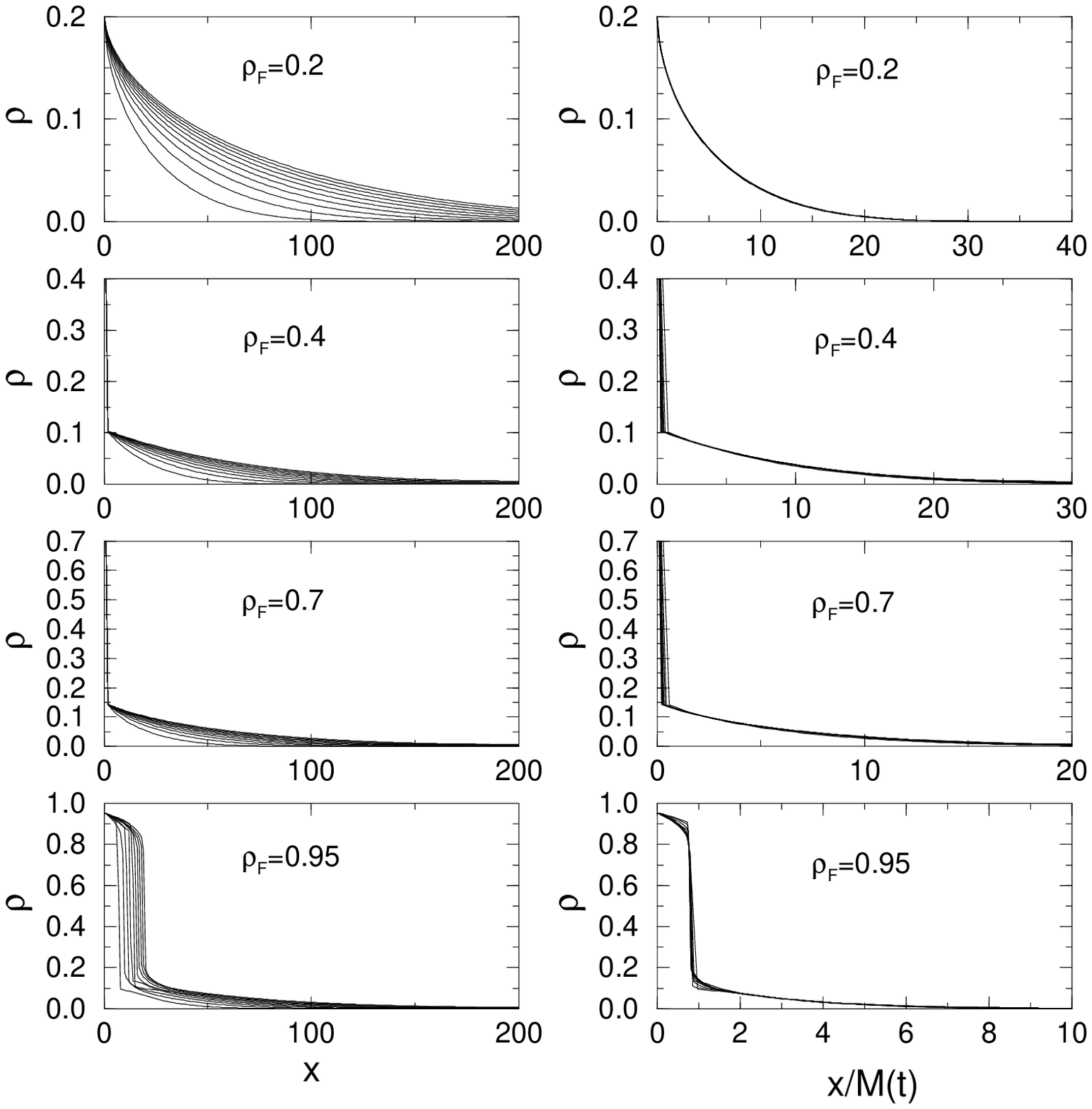}
\caption{\label{Fig4} {\small  Dynamical density profiles defined by Eq.(\ref{m}) for $\beta J = 3$ and different values of
$\rho_F$. In the left column $\rho_t(X)$ is plotted versus the space variable $X$ and here different curves on each graph show
the dynamics of the density distribution in the single-file pore, Eq.(\ref{desired1}), while in the right column we present corresponding plots 
for $\rho_t(X)$ vs the scaling variable $X/M(t)$.}}
\end{center}
\end{figure} 

We begin with a small $\beta J$ limit, in which case some perturbative analytical
calculations are possible. To do this, let us represent 
$\rho(\omega)$ in the form of the series
\begin{equation}
\label{series}
\rho(\omega) = \sum_{n=0}^{\infty} (2 \beta J)^n \rho_n(\omega), 
\end{equation}
(which is related to expansion in powers of the reservoir
density $\rho_F < 1$, as we will see in what follows) and try to calculate explicitly
several first terms in such an expansion,  constraining ourselves
to the quadratic in $\beta J$ approximation.  After some rather cumbersome but straightforward calculations, we find
 eventually, that 
in the quadratic with respect to the parameter $2
\beta J$ approximation, the prefactor $m(J, \rho_F)$ obeys:

\begin{eqnarray}
\label{mm}
m(J,\rho_F) &=& \frac{\rho_F }{\sqrt{\pi}}  \; - \; (2 \; \beta \; J) \; \Big[0.18 \;
 \rho_F^2  \; 
- \; 0.134 \;
 \rho_F^3\Big] \; \nonumber\\
& -& \; (2 \; \beta \; J)^2 \; \Big[0.025 \;  \rho_F^3 \;
- \; 0.047 \;   \rho_F^4 \; + \; 0.018 \;  \rho_F^5\Big]
\end{eqnarray}

This dependence, as one notices, 
agrees quite well with the numerical solution for relatively low values of $\beta J$ over the entire domain of variation of
$\rho_F$, or, for
larger $\beta J$, for progressively smaller values of $\rho_F$. 

To get some understanding of the intricate, 
non-trivial behavior of $m(J,\rho_F)$, observed for larger values of
$\beta J$, we analyse dynamical density profiles defined by Eq.(\ref{m}) versus
$X$ and 
$X/M(t)$
for $\beta J = 3$ and different reservoir densities $\rho_F$, (see Fig.4). 
Note now that the form of the density profiles is rather complex and depends largely
on whether $\rho_F$ is less than or exceeds $\rho_{c,\pm}$.  
When $\rho_F \leq \rho_{c,-}$, the
form of the density profile is well described by  $\rho_t(X) = \rho_F {\rm erfc(X/M(t))}$,
where ${\rm erfc(X/M(t))}$ is the error function. This behavior is essentially the
same as the one predicted for non-interacting lattice gas in \cite{bur}. On the other
hand, when  $\rho_F$ exceeds $\rho_{c,-}$, but is less than $\rho_{c,+}$ (for $\beta J = 3$, $\rho_{c,\pm}$ are equal to $0.79$
and $0.21$, respectively)
we have two different regimes: the density rapidly, within the small constant distance $l(\rho_F)$,
 drops to the value $\rho_{c,-}$ and then evolves as $\rho_t(X) = \rho_{c,-} {\rm erfc(X/M(t))}$, where
$\rho_{c,-}$ is independent
of $\rho_F$.
Since,  the prefactor $m(J,\rho_F)$ is just an integral of $\rho_t(X)$, 
we have, hence, two contributions: the first is the integral over the interval $[0,l(\rho_F)]$, which is weakly dependent on
$\rho_F$ and the second one - the integral over $[l(\rho_F),\infty[$, which is independent of $\rho_F$ and gives the bulk
contribution to $m(J,\rho_F)$. These two contributions
define the plateau-like part in the dependence of $m(J,\rho_F)$ on $\rho_F$. 
On the other hand, when  $\rho_F$ exceeds $\rho_{c,+}$ and
hence $D(\rho)$ is positive definite, 
a dense droplet-like
structure emerges near the entrance of the pore, 
which grows in size in proportion to $\sqrt{t}$ and 
contains most of the particles. This phenomenon 
explains apparently the growth of  
$m(J,\rho_F)$ with $\rho_F$ observed for $\rho_F \geq \rho_{c,+}$.

\vspace{0.3in}

{\Large \bf Conclusions}

\vspace{0.2in} 

In conclusion, we have studied propagation dynamics of a particle phase emerging in a
single-file pore connected to a reservoir of particles (bulk liquid phase). Modelling
the pore  as a semi-infinite
one-dimensional regular lattice, whose sites support, at most, 
a single occupancy, and
supposing that 
the particles interaction potential consists of an
abrupt, hard-core repulsive part, and is attractive, with an amplitude $J \geq 0$,  
for the nearest-neighboring particles,  we derived the dynamical equation
describing the time-evolution for the local particle density. This equation has
been 
analysed both analytically and
numerically. Within our appoach, we defined the evolution of the total number $M(t)$ 
of particles entering 
the single-file pore up to time
$t$ and also discussed the dynamics of the density distribution function within the
 pore. We have shown that the growth of 
$M(t)$ is described by $M(t) = 2 m(J,\rho_F) \sqrt{D_0 t}$, which can be thought of
 as the microscopic analog of the Washburn
equation, where $D_0$ denotes the "bare" diffusion coefficient, while the prefactor 
$m(J,\rho_F)$ is a non-trivial function of the attractive
interactions amplitude $J$ and the density $\rho_F$ at the entrance to the pore. We have discussed
a peculiar behavior of $m(J,\rho_F)$ and explained it through the analysis of the dynamical density distribution
within the pore.

\vspace{0.2in}

\renewcommand{\baselinestretch}{1.0}

\end{document}